\documentclass{elsart}
\usepackage{epsfig}
\newcommand{\beq}{\begin{equation}}
\newcommand{\eeq}{\end{equation}}
\newcommand{\eq}[1]{(\ref{#1})}

\begin{document}

\begin{frontmatter}

\title{Delbr\"uck scattering and the $g$-factor of a bound
electron}

\author[vniim,mpq]{Savely G. Karshenboim}
\ead{sek@mpq.mpg.de}
\and
\author[binp]{Alexander I. Milstein}
\ead{milstein@inp.nsk.su}
\address[vniim]{D. I. Mendeleev Institute for Metrology (VNIIM),
 St. Petersburg 198005, Russia}
\address[mpq]{Max-Planck-Institut f\"ur Quantenoptik, 85748 Garching, Germany}
\address[binp]{Budker Institute of Nuclear Physics, 630090 Novosibirsk,  Russia}

\begin{abstract}
The leading contribution of the light-by-light scattering effects
to $g$-factor of a bound electron is derived. The corresponding
amplitude is expressed in terms of low-energy Delbr\"uck
scattering of a virtual photon. The result reads $\Delta g=
(7/216)\,\alpha(Z\alpha)^5$.
\end{abstract}

\begin{keyword}
Quantum electrodynamics \sep $g$ factor \sep Delbr\"uck scattering
\PACS 12.20.-m \sep 31.30.Jv \sep 32.10.Dk
\end{keyword}
\end{frontmatter}

\section{Introduction}

Recently a significant progress in measurements of the $g$-factor
of a bound electron in a hydrogen-like ion with a spinless nucleus
has been achieved \cite{carbon,oxigen}. At the same time an
accurate theory was successfully developed
\cite{beier,pla,h2,jetp,prl,recoil1,recoil2,oneloop}. Comparison
of theory and experiment \cite{carbon} for the hydrogen-like
carbon ion $^{12}{\rm C}^{5+}$ allows to determine the most
accurate values of the electron mass \cite{prl}
\begin{equation}\label{meu}
m_e = 0.000\,548\,579\,909\,2(4)~{\rm u}
\end{equation}
and the proton-to-electron mass ratio
\begin{equation}\label{mpme}
m_p/m_e = 1836.152\,673\,3(14)\;.
\end{equation}
Those are three times more accurate than the recommended CODATA
values \cite{codata} based on study of protons and electrons in
Penning trap \cite{farnham}.

\begin{figure} 
\begin{minipage}[b]{0.46\textwidth}
\centerline{\epsfbox{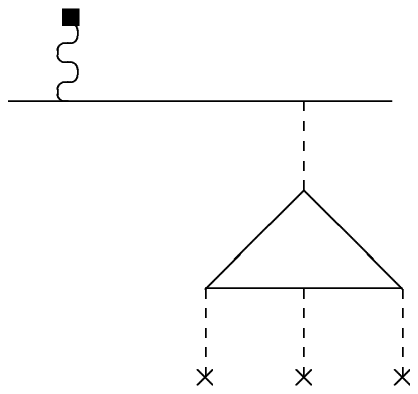}}
\end{minipage}%
\hskip 0.08\textwidth
\begin{minipage}[b]{0.46\textwidth}
\centerline{\epsfbox{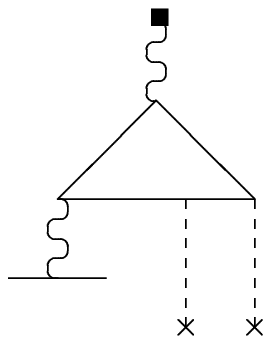}}
\end{minipage}
\vspace{10pt}
\begin{minipage}[t]{0.46\textwidth}
\caption{\label{fig1} One of the diagrams for the ``electric loop"
contribution to the $g$-factor of a bound electron. The internal
dashed line corresponds to $D^{00}$ component of the photon
propagator. The dashed line with the cross denotes a Coulomb
field, the wavy line with the square denotes the external
homogeneous magnetic field.}
\end{minipage}%
\hskip 0.08\textwidth%
\begin{minipage}[t]{0.46\textwidth}
\caption{\label{fig2} One of the diagrams for the ``magnetic loop"
contribution to the $g$-factor of a bound electron. The internal
wavy line corresponds to $D^{ij}$ component of the photon
propagator. Other notations are the same as in Fig.\ref{fig1} }
\end{minipage}
\end{figure}

Theory and experiment equally contribute into uncertainty of the
values in Eqs. \eq{meu} and \eq{mpme}. An essential part of
theoretical uncertainty (maybe even the dominant part) is due to
the light-by-light scattering effects. A part of them is related
to the Wichmann-Kroll potential (so-called the ``electric-loop''
term presented in Fig.~1, where the contribution of a vacuum
polarization by free electrons known analytically for a point-like
nucleus \cite{jetp} is not included). Its leading contribution
($\propto \alpha(Z \alpha )^6$)  has been found analytically
\cite{pla}: \beq \label{gEL} \Delta g({\rm EL}) =
2\left(\frac{38}{45}-\frac{2\pi^2}{27}\right) \frac{\alpha(Z
\alpha )^6}{\pi}\,. \eeq Here $Z$ is the nuclear charge number,
$\alpha=e^2$ is the fine-structure constant, $\hbar=c=1$. The
other part (so-called the ``magnetic loop'' term) presented in
Fig.~2 has not been known. Even the order of magnitude of the
leading term has not been clarified. Some rough estimations for it
were included into evaluations in Ref.~\cite{beier,jetp,prl}. We
claim that the magnetic-loop effects contribute in order
$\alpha(Z\alpha)^5$. Therefore, potentially they could strongly
affect the $g$ factor of a bound electron. For instance, with
\[
{\Delta g({\rm ML} )\over 2} = C_{\rm ML} \,\alpha(Z\alpha)^5
\]
one can find a shift of $g$-factor up to $2C_{\rm ML} \cdot
10^{-9}$ in the case of the hydrogen-like ion of carbon,  and
$C_{\rm ML} \cdot 10^{-8}$ in the case of oxygen. In this paper we
study the leading contribution of the magnetic-loop effects and
determine coefficient $C_{\rm ML}$.

\section{Low-energy Delbr\"uck scattering and the contribution of ``magnetic loop''}

First we note that we need to study a vacuum polarization loop
with the two lines of external virtual photons in the Coulomb
field of a nucleus. One of these photons is related to an
homogenous magnetic field and the other  connects the electron
loop and the atomic electron line. Because of the Furry theorem
the leading effects of the electric field are related to the
diagram with the two Coulomb lines. Then, we apply the block of
the vacuum polarization  when both external photon lines transfer
momenta ${\bf k}_1$ and ${\bf k}_2$ significantly smaller than the
electron mass,  $\vert{\bf k}_2\vert \sim Z\alpha m_e$. The
rigorous analysis shows that the contribution to g-factor of the
region $\vert{\bf k}_2\vert\sim m_e$ is  of order of
$\alpha(Z\alpha)^6$. The kinematics $k_1,\,k_2\ll m_e$ is similar
to that for low-energy  Delbr\"uck scattering (see  Fig.~3).

\begin{figure}[h]
\epsfxsize=5cm \centerline{\epsfbox{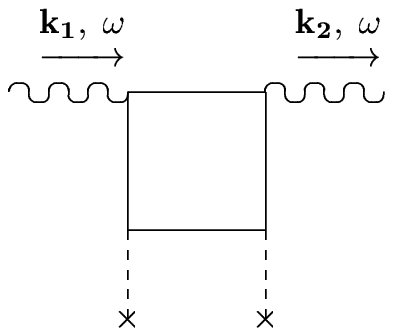}}
\caption{\label{fig3} A diagram for Delbr\"uck scattering. The wavy line
indicates here the incoming or outgoing photon.}
\end{figure}

Delbr\"uck scattering (see \cite{delb1,delb2}) is a process in
which the initial photon turns into a virtual electron-positron
pair that scatters in the electric field of an atom and then
transforms into the final photon. The Feynman diagram
corresponding to the amplitude in the lowest in $Z\alpha$ order is
shown in Fig.~3. Let us consider the scattering of a virtual
photon with the initial momentum $k_{1}=(\omega, {\bf k}_{1})$ and
the final momentum $k_{2}=(\omega, {\bf k}_{2})$, with
$\omega,\,|{\bf k}|_{1,2}\ll m_e$. Due to the gauge invariance,
the Delbr\"uck scattering amplitude reads
\begin{eqnarray}\label{Tmn}
T^{\mu\nu}&=&\frac{\alpha(Z\alpha)^2}{m_e^3}\{C_1\cdot[g^{\mu\nu}(k_1\cdot
k_2)-k_2^\mu k_1^\nu]\nonumber\\
&~& + C_2\cdot[\omega^2g^{\mu\nu}-\omega(n^\mu k_1^\nu+k_2^\mu
n^\nu)+(k_1\cdot k_2)n^\mu n^\nu]\}\, ,
\end{eqnarray}
where  $n^\mu=g^{\mu \,0}$, $C_1$ and $C_2$ are some constants. This
form of the amplitude follows from  the relations
$k_{1\mu}T^{\mu\nu}=T^{\nu\mu}k_{2\mu}=0$, and from the linearity
of $T^{\mu\nu}$ with respect to $k_{1}$ and $k_{2}$. To
calculate the magnetic loop contribution we need only the
amplitude $T^{ij}$ at $\omega=0$. We can find it since the
amplitude in Eq.\eq{Tmn} was derived for the real photons. For
$\omega=|{\bf k}|_1=|{\bf k}|_2$  the amplitude
$T_D=e_\mu^{(1)}T^{\mu\nu}e_\nu^{(2)*}$ ($e_\mu^{(1,2)}$ are the
polarization vectors) is of the form
\begin{eqnarray}\label{real}
T_D=\frac{\alpha(Z\alpha)^2}{m_e^3}\{-(C_1+C_2)\omega^2({\bf
e}_1\cdot{\bf e}_2^*)+ C_1[{\bf e}_1\times {\bf k}_1][{\bf
e}_2^*\times {\bf k}_2]\}\, .
\end{eqnarray}
The amplitude $T_D$ (\ref{real}) was obtained in Ref.~\cite{CTP}
(see also \cite{PM95}). Using the results of \cite{CTP} and
Eq.(\ref{real}), we obtain
\begin{eqnarray}\label{AB}
C_1= {7\over 16\cdot72}\quad\quad {\rm and} \quad\quad C_2=-{73\over
32\cdot72}\,.
\end{eqnarray}
Let us consider now the amplitude $T_M$  of interaction of the
magnetic field with the spin part of the magnetic moment of the
atomic electron. In the zero approximation it reads
\begin{eqnarray}\label{T0}
T_M^{(0)}&=& -{ie\over m_e}\int \frac{d^3{\bf k}}{(2\pi)^3}\,
 {\bf A_k}\cdot[{\bf k}\times \bf{s}] \,\rho_{\bf k}\nonumber\\
 &=& {e\over m_e}\int \frac{d^3{\bf k}}{(2\pi)^3}\,
 {\bf B_k}\cdot \bf{s} \,\rho_{\bf k}
 \; ,
\end{eqnarray}
where ${\bf A}_k$ is the vector potential, $\bf{s}$ is the spin
operator, and $\rho_{\bf k}$ is the Fourier transform of the
electron density $|\psi({\bf r})|^2$, $\psi({\bf r})$ is the
bound electron wave function. Note that a sign of $T_M$ is
opposite to that of Hamiltonian.  In the case of the homogeneous
magnetic field ${\bf B}$, we have
\[
{\bf B_k}=i[{\bf k}\times {\bf A_k}] = (2\pi)^3\,\delta({\bf k})\,{\bf
B} \;,
\]
and we can replace in Eq.~(\ref{T0}) $\rho_{\bf k}$ by
$\rho_{0}=1$. Using (\ref{Tmn}) we can represent the correction
$T_M^{(1)}$ as follows:
\begin{eqnarray}\label{T1}
T_M^{(1)}&=& -\frac{ie\alpha(Z\alpha)^2}{m_e^4}\,C_1\int \frac{d^3{\bf
k}}{(2\pi)^3}\int \frac{d^3{\bf q}}{(2\pi)^3}\,{4\pi\over {\bf q}^2}\nonumber\\
&~&\quad\quad\times
\biggl\{({\bf k}\cdot{\bf q}){\bf A}_k-({\bf A}_k\cdot{\bf q}){\bf k}\biggr\}\cdot
 [{\bf q}\times \bf{s}] \,\rho_{\bf q}\nonumber\\
&=&-\frac{ie\alpha(Z\alpha)^2}{m_e^4}\,C_1\int \frac{d^3{\bf
k}}{(2\pi)^3}{\bf A_k}
  \cdot[{\bf k}\times {\bf s}]
\int \frac{d^3{\bf q}}{(2\pi)^3}\,{8\pi\over {3}}\,\rho_{\bf q}\nonumber\\
&=&-\frac{ie\alpha(Z\alpha)^2}{m_e^4}\,C_1\int \frac{d^3{\bf
k}}{(2\pi)^3}{\bf A_k} \cdot[{\bf k}\times {\bf s}]\cdot
\frac{8\pi}{3}|\psi(0)|^2 \quad .
\end{eqnarray}
Thus, our final result for the relative correction of the magnetic
loop reads
\begin{eqnarray}\label{final1}
\frac{\Delta g({\rm ML})}{2}=T_M^{(1)}/T_M^{(0)}= \frac{7\pi
\alpha(Z\alpha)^2}{6\cdot 72\,m_e^3}\,|\psi (0)|^2 \; .
\end{eqnarray}
 Substituting $|\psi (0)|^2=(Z\alpha m)^3/\pi$ for $1s$ state, we
 have
\begin{eqnarray}\label{final}
\Delta g({\rm ML})= \frac{7 }{216}\alpha(Z\alpha)^5  \; ,
\end{eqnarray}
For the interesting cases of hydrogen-like carbon ($Z=6$)
and oxygen  ($Z=8$) the numerical values are $0.4\cdot 10^{-10}$
and $1.6\cdot 10^{-10}$, respectively. The obtained correction has the same
order of magnitude as results in ref.~\cite{beier}. However,
 a quantitative comparison of our analytic result with numerical
 calculations there is not helpful because of lack of accurate
 numerical data for $Z\leq10$.

\section{Conclusions}

If one applies Eq. (\ref{final}) to the $g$-factor for
hydrogen-like carbon and oxygen, then the result in some sense is
controversial. First,  it has  lower order of magnitude than
expected, i.e. $(\alpha(Z\alpha)^5)$. On the other side, the
correction \eq{final} is consistent with the preliminary estimate
(see, e.g., \cite{jetp}) because of its very small numerical
coefficient. Next, it is not clear if the leading in $Z\alpha$
term gives a dominant contribution. E.g., in the case of recoil
correction for carbon, the leading term is smaller than the
next-to-leading term because of the small coefficients in it
\cite{recoil1,recoil2}.

To estimate uncertainty related to the higher order terms, we note
that the coefficients in the contribution of the free vacuum
polarization \cite{jetp}
\begin{eqnarray}
\Delta g({\rm VP}) &=& 2\cdot
 \frac{\alpha}{\pi}\cdot \left[ - \frac{8}{15} (Z\alpha)^4
  + \frac{5\pi}{18} (Z\alpha)^5  \right]\nonumber\\
  &\simeq& -0.34 \times \alpha(Z\alpha)^5
  +0.56 \times\alpha(Z\alpha )^6
\end{eqnarray}
are about unity in contrast to the light-by-light contributions
 \eq{gEL} and \eq{final},
\begin{eqnarray}
\Delta g({\rm ML}) &\simeq& 0.032 \times
\alpha(Z\alpha)^5\;,\nonumber\\
\Delta g({\rm EL}) &\simeq& 0.072\times\alpha(Z \alpha )^6 \;.
\end{eqnarray}
We consider an estimation $\pm \alpha(Z \alpha)^6$ for higher order magnetic-loop
 effects as a conservative one.
  It is below $10^{-10}$ at $Z<6$, leads to $\Delta g \sim 1\cdot 10^{-10}$
   for carbon and $\Delta g\sim 6\cdot 10^{-10}$ for oxygen. Our results are
collected in the Table~1.

\begin{table}
\label{table1}
\begin{center}
\begin{tabular}{ccc}
\hline
Ion & $g$ & Ref.$^*$ \\
\hline
$^4$He$^+$ & 2.002\,177\,406\,7(1) & \protect\cite{kinew} \\
$^{10}$Be$^{3+}$ & 2.001\,751\,574\,5(4)  & \protect\cite{kinew} \\
$^{12}$C$^{5+}$ & 2.001\,041\,590\,1(4)& \protect\cite{oneloop} \\
$^{16}$O$^{7+}$ & 2.000\,047\,020\,1(8)& \protect\cite{oneloop} \\
\hline
\end{tabular}
\end{center}
\caption{The bound electron $g$ factor in low-$Z$ hydrogen-like
ions with spinless nucleus. The uncertainty for the two-loop
contribution is taken from [6]. Ref.$^*$ is related to the
one-loop result for the self-energy contribution. For lighter
atoms it is taken from [17] based on fitting data of [3], while
for heavier isotopes we use the
 results of [10].}
\end{table}

The experimental results for $g$ are available for the ions of
carbon and oxygen  with a relative uncertainty of $2\cdot10^{-9}$
being limited by our knowledge of the electron mass
\cite{farnham}. However, their ratio is free of this uncertainty
\beq g(^{12}{\rm C} ^{5+}) /g(^{16}{\rm O} ^{7+})
=1.000\,497\,273\,1(15)\,, 
\eeq
 and  is in a fair agreement with
the theoretical prediction 
\beq 
g(^{12}{\rm C}^{5+}) /g(^{16}{\rm O}^{7+}) =1.000\,497\,273\,3(3) \,. 
\eeq
Calculation of the next-to-leading term of the magnetic-loop effects is in progress.

\section*{Acknowledgements}

We are grateful to Peter Mohr, G\"unther Werth and Vladimir Ivanov
for stimulating discussions. We  thank the School of Physics at
the University of New South Wales for their warm hospitality during the stay when
this work has been done. The work of S.G.K. was supported in part
by RFBR grant 00-02-16718, and the work of A.I.M. by UNSW.

\end{document}